\documentstyle[multicol,epsf,aps,amsmath,prl]{revtex}
\begin{document}
\draft
 
\title{Quantum entanglement in carbon nanotubes}

\author{Cristina Bena$^1$, Smitha Vishveshwara$^1$, Leon Balents$^1$,
Matthew P. A. Fisher$^2$}
\address{$^1$Department of Physics, University of California, 
Santa Barbara, CA 93106 \\
$^2$Institute for Theoretical Physics, University of California,
Santa Barbara, CA 93106--4030
}

\date{\today}
\maketitle

\begin{abstract}
   With the surge of research in quantum information, the issue of
  producing entangled states has gained prominence.  Here, we show
  that judiciously bringing together two systems of strongly
  interacting electrons with vastly differing ground states - the
  gapped BCS superconductor and the Luttinger liquid, - can result in
  quantum entanglement. We propose three sets of measurements
  involving single-walled metallic carbon nanotubes (SWNT) which have
  been shown to exhibit Luttinger liquid physics, to
  test our claim and as nanoscience experiments of interest in and of
  themselves.
\end{abstract}


\begin{multicols}{2}
\narrowtext 
Entangled pairs are quantum entities consisting of two components sharing
a common wave function; a measurement on one component predetermines
the state of the other \cite{EPR}.  Such pairs are a basic resource
for quantum information processing, and recent years have begun to see
many promising approaches to their production in small numbers (e.g.
\cite{entangle1}).  For large scale implementation of
quantum information technology, a realization of entanglement in solid
state systems and an appropriate means of transporting the components
of the entangled pair over long distances are essential.  While the
constituent Cooper pairs of a gapped BCS superconductor have been
studied as a possible natural source of such pairs
\cite{loss}, the question of separating and transporting them
requires further investigation. Here we show how electron-electron
interactions in one dimension enables sequential injection of entangled
pairs from a superconductor into {\it two} nanotubes.  The
SWNTs, in turn, would allow for transport of entangled states over
appreciable distances.

SWNTs, essentially long conducting cylinders of nanoscale diameters
and lengths '$L$' of several microns, are indeed well-suited as carriers
for coherent spin states. They are extremely pure systems with large
Fermi velocities of $v_F \sim 10^6 m/s$, and are known to exhibit
ballistic transport over very long distances.  In particular, at low
energies compared to the subband spacing $\epsilon_0 \sim 1eV$,
transport is characterized by four ballistic modes propagating with
linear dispersion.  At low temperatures, $T \lesssim T_\phi=\hbar
v_F/k_B L$, electrons can thus travel the entire length of the tube
without losing coherence due to thermal effects.  Moreover, nanotubes are
expected to be nearly ideal spin conductors\cite{BalentsEgger},
as indeed supported by recent experiments\cite{alphenaar}. In
addition, they can be grown in a controlled fashion, and as
illustrated by their recent employment in building electronic circuit
elements\cite{dekker}, they show a manipulability fit for
solid state devices.  While all these features bode well for transport
and usage of entangled pairs, as detailed in what follows, the actual
injection and separation of these pairs into the two tubes rely on the
fact that nanotubes have demonstrated Luttinger liquid behavior
characteristic of electrons interacting in
1D\cite{nanexp2}.  

\begin{figure}[h]
\epsfxsize=2.5in
\centerline{\epsffile{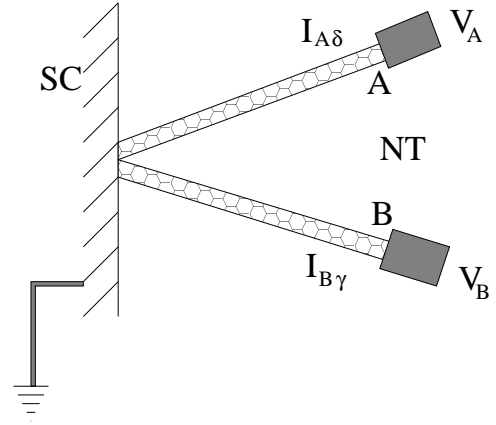}}
\vspace{0.15in}
\caption{\label{split} Set-up of two nanotubes A and B end-contacted
  to a superconductor.  
Voltage drops $V_A$ and $V_B$ may be preferentially 
applied across tubes A and B respectively, and 
currents through each of them may be measured.}
\vspace{0.15in}
\end{figure} 
\vspace*{-.1in}
The basic set-up we consider consists 
of two nanotubes A and B, end-contacted well
within a coherence length of each other to a gapped singlet-paired 
superconductor as in Fig.1. Each wire
can be described in bosonized language 
by a four channel Luttinger liquid Hamiltonian \cite{Klm,egger}
\begin{equation}
H_{i}= \sum_{a}\int_0^{\infty}\! dx v_{a}
[g_{a}^{-1}(\partial_x\theta^i_{a})^2 + g_{a}
 (\partial_x\phi^i_{a})^2],
\label{ham}
\end{equation}
where $i=A,B$ denotes the wires, and
$a=\rho_{\pm}, \sigma_{\pm}$ correspond to the four free sectors 
of the theory where, by linear transformations we have 
made a change of
basis from the spin-channel indices ($1\!\uparrow,1\!\downarrow, 
2\!\uparrow,2\!\downarrow$). The relation between the bosonic fields
$\theta^i_{n \alpha}$,
$\phi^i_{n \alpha}$ ($n=1/2, \alpha=\uparrow/\downarrow$) and
the original chiral right-/left-moving 
fermionic fields $\psi^i_{R/L n \alpha}$ 
is expressed through
the Bosonization procedure via the transformation
$\psi^i_{R/L n \alpha} \sim e^{i(\phi^i_{n \alpha} \pm \theta^i_{n \alpha})}$. 
The parameter
$g_{a}$ captures the strength of interactions; $g_{a}=1$ for the 
non-interacting channels $a=(\rho-,\sigma\pm)$, while for the charge sector
$g_{\rho+}$ has
the value $g_{\rho+}=g \approx 0.25 $\cite{nanexp2,Klm}.
The velocities of the free modes $v_a$ are given by $v_a=v_F/g_a$.

As it is desirable to inject entangled pairs individually, we
focus on the case of high resistance contacts where successive Cooper pairs
hop sequentially from or into the superconductor. This limit corresponds
to almost perfect backscattering at the superconductor-nanotube interface, 
whence
$\psi^i_{n L \alpha}(0)=\psi^i_{n R \alpha}(0)$ where $i=A,B$ denotes
the wire and $n$ refers to
the channel indices $1$ and $2$.
In the bosonized language,
these boundary conditions become
$\theta^i_{a}(0)=0$ for $i=A,B$ and all  $a=\rho_{\pm}, \sigma_{\pm}$.

In this set-up we analyze perturbatively the effects of a small amount of
Cooper pair tunneling between the superconductor and the two wires. 
The corresponding Hamiltonian $H_t$ for such processes is  given by
$H_t=\sum_{i=A,B} H^t_{ii} +H^t_{AB}$ where $H^t_{ii}$ describes the tunneling
of whole pairs into wire `$i$', and $H^t_{AB}$ describes 
processes in which one electron 
of the pair tunnels into the tube A and the other one into the tube $B$.
Thus
\begin{eqnarray}
H^t_{ii}(x)=&& v_F T_{ii} [\psi^{\dagger}_{i\uparrow b}(0)
\psi^{\dagger}_{i\downarrow c}(0)\nonumber \\&&
-\psi^{\dagger}_{i\downarrow b}(0)
\psi^{\dagger}_{i\uparrow c}(0)+ (h.c.)], 
\nonumber
\\
H^t_{AB}(x)=&&v_F T_{AB}[\psi^{\dagger}_{A\uparrow b}(0)
\psi^{\dagger}_{B\downarrow c}(0)\nonumber \\&& 
- \psi^{\dagger}_{A\downarrow b}(0)
\psi^{\dagger}_{B\uparrow c}(0) + (h.c.)],  
\label{tunn}
\end{eqnarray}
where $\psi^{\dagger}_{i\sigma b}$ are
creation operators for electrons in wire `$i$' with corresponding
spin $\sigma$ and flavor $b=1R,1L,2R$ and $2L$. 
The coefficients $T_{ii}$ and $T_{AB}$ are the bare tunneling amplitudes.

As the common
wave-function for each Cooper pair in the bulk is peaked at its center
of mass, in a non-interacting model the transmission probability for a
pair to enter e.g. wire A, $t_{AA} \propto |T_{AA}|^2$, is much higher than the
transmission probability for the pair-splitting process, $t_{AB} \propto
|T_{AB}|^2$, i.e. $t_{AA} \gg t_{AB}$.  
Naively, then, one might expect each Cooper pair
to tunnel entirely into one tube or the other.  However, due to
interactions, tunneling of charge into the ends of the nanotubes
involves more than the mere overlap of single-particle electronic
wave-functions between the tube and the superconductor.  Addition of
one extra electron into a tube involves the coherent rearrangement of all
electrons in its bulk.  As a consequence of this Luttinger
liquid physics, one can show that the single electron tunneling
density of states at a low energy $E$ compared to $\epsilon_0$ goes as
$\rho_e(E) \sim
\epsilon_0^{-1}(E/\epsilon_0)^{\frac{1}{4}(\frac{1}{g}-1)}$, while
the density of states available to tunnel in a Cooper pair is
$\rho_{2e}(E) \sim
\epsilon_0^{-1}(E/\epsilon_0)^{\frac{1}{g}}$. 
If the two
nanotubes are raised to a voltage, $V \ll k_B T/e$ above the
superconductor, Fermi's Golden Rule reveals that the rate
$\Gamma_{AA}$, at which entire Cooper pairs tunnel from the
superconductor into the end of one tube, is proportional to $eV
\rho_{2e}$, and at any given temperature $T$, has the dependence
$\Gamma_{AA} \sim (eV/h)(k_B T/\epsilon_0)^{\frac{1}{g}-1}$.  However, the rate
$\Gamma_{AB}$, at which split entangled pairs are injected into
both tubes is proportional to the convolution of their one particle
tunneling densities of states and has the dependence $\Gamma_{AB} \sim
(eV/h) (k_B T/ \epsilon_0)^{\frac{1}{2}(\frac{1}{g}-1)}$.
Remarkably, at low temperature, $\Gamma_{AB} \gg \Gamma_{AA}$, and thus
almost all charge transfer occurs as split entangled pairs.  We now
turn to three sets of measurements that capture these principles in a
concrete manner.

The simplest experimental signature of the splitting of Cooper pairs
may be obtained from the transconductance measured for two nanotubes
as shown in Fig.1. 
In response to a voltage difference between the nanotubes and 
the superconductor, 
we compute the resulting currents flowing into the two wires.
We start from the Hamiltonian of Eq. (\ref{ham})
with the appropriate boundary conditions and, along the lines of
Ref. \cite{Keldysh2},
we use a non-equilibrium Keldysh technique
\cite{Keldysh1} 
perturbative in the amount of Cooper pair tunneling described by 
Eq. (\ref{tunn}). 
To bring out the physics of the Cooper pair splitting, we first
consider the specific case of applying a voltage drop $V_A$ across tube
A and none across tube B under the condition 
$k_B T \ll e V_A \lesssim \epsilon_0, \Delta$, where $\Delta$ 
is the superconducting gap. 
In tube A, the applied voltage 
would produce a current with two components -- one due to entire
pairs tunneling in A, $I_{AA} \sim t_{AA} [(2 e)^2 /h] 
(e V_A /\epsilon_0)^{2 \alpha} V_A$, and another due to the splitting
of pairs into the two tubes,  $I_{AB} \sim t_{AB} (e^2 /h)
(e V_A /\epsilon_0)^{\alpha} V_A$, where $\alpha=(1/g-1)/2
\approx 1.5$.  
Strikingly, the current $I_{AB}$ runs equally in both
tubes, in spite of the absence of a voltage drop 
across tube B.  
The non-linear behavior of current is a reflection of the power laws in
the density of states, and despite the fact that
$t_{AB}<t_{AA}$, the contribution from the
pair split process clearly dominates at low voltages. 

More generally, at finite
temperature $T$, when voltage drops $V_A$ and $V_B$ are present across
both tubes, one can define an associated conductance matrix
$G_{ij}=\partial I_i/\partial V_j$, where $i,j$ stand for the
tubes A and B.  It has the form
\begin{eqnarray}
G_{AA}&&= t_{AA} \frac{(2e)^2}{h}\Big[\frac{k_B
  T}{\epsilon_0}\Big]^{2\alpha} {\cal F}_{2\alpha}\Big[\frac{2e
  V_A}{k_B T}\Big]  
\nonumber \\ &&
  + t_{AB} \frac{e^2}{h} \Big[\frac{k_B
  T}{\epsilon_0}\Big]^{\alpha} {\cal F}_\alpha \Big[\frac{e
  (V_A+V_B)}{k_B T}\Big], 
\label{condmat0}
\end{eqnarray}
\begin{eqnarray}
G_{AB}&&= t_{AB}  \frac{e^2}{h} \Big[\frac{k_B
  T}{\epsilon_0}\Big]^{\alpha} {\cal F}_\alpha \Big[\frac{e
  (V_A+V_B)}{k_B T}\Big], \label{condmat}
\end{eqnarray}
with $G_{BB}$ obtained from $G_{AA}$ by interchanging $A$ and $B$.
Here, the scaling function ${\cal F}_\alpha[x] =
\partial_x[2 \sinh (x/2) |\Gamma(1+\alpha/2 + i x/2\pi)|^2]$ (where
$\Gamma(z)$ is the gamma function) and has the limits ${\cal
  F}_\alpha(x) \rightarrow_{x\rightarrow 0} |\Gamma(1+\alpha/2)|^2$, ${\cal
  F}_{\alpha}(x) \rightarrow_{x\rightarrow\infty} (1+\alpha)
(x/2\pi)^{\alpha}$.  The dominance of split Cooper pair injection in
charge transfer is directly seen in Eqs.~(\ref{condmat0},\ref{condmat}), 
which implies that for $k_B T \ll \epsilon_0$, the two differential
conductances $G_{AA} \approx G_{AB}$.

The transconductance experiment directly measures the fact that
charge is simultaneously injected into both nanotubes.  It is not,
however, sensitive to the spin state of these electrons.  We next
consider a Josephson current measurement which verifies that spin
$1/2$ is added to each wire. Here, as in
Fig.2, we consider two nanotubes A and B, meeting at points X
and Y separated by distance '$L$' along the tubes. 
\begin{figure}
\epsfxsize=2.5in
\centerline{\epsffile{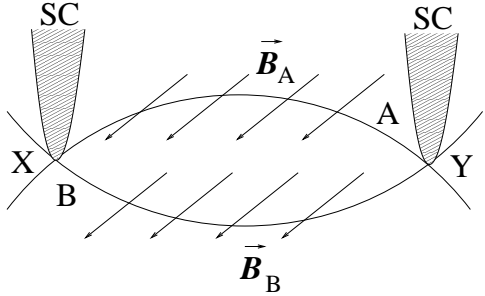}}
\vspace{0.15in}
\caption{Set-up of two infinite nanotubes A and B crossing at points X and Y 
separated by a distance 'L' along each tube. Superconducting point contacts at
junctions X and Y allow for Cooper pair tunneling into the tubes.}
\vspace{0.15in}
\label{jj}
\end{figure} 
\vspace*{-0.1truein}
At each junction,
a superconducting lead makes a point contact with both tubes.  To probe the
spin state of the electrons, tube A (tube B) is subjected to a magnetic field
 $\vec{B}=B_A \hat{x}$ ($\vec{B}=B_B \hat{x}$).  It may be convenient
 to choose the field axis $\hat{x}$ parallel to the junction so as to
 minimize orbital effects.
For this particular set-up the Hamiltonian for each wire is similar to
the one described in Eq. (\ref{ham}), with the limits of integration extending
over the entire range in the position space, as opposed to the 
imposition of hard boundary conditions of the previous set-up. 
We compute the free energy of the system perturbatively in the tunneling of
Cooper pairs at points X and Y described by a tunneling term
similar (but not identical) to Eq. (\ref{tunn}). We find that,
at low temperature (when the thermal coherence length $L_\phi = \hbar
v_F/k_B T > L$), the nanotubes act as a Josephson weak link with associated 
critical current
\begin{eqnarray}
&&I_c =\frac{e v_F}{L} \bigg\{\sum_{i=A,B}\left[
\tilde{t}_{ii}^{(1)} 
\Big(\frac{d}{L}\Big)^{2 \beta} +
\tilde{t}_{ii}^{(2)} \Big(\frac{d}{L}\Big)^{\alpha}
\cos\Big({\frac{{\bf
        g} \mu_B B_i L}{\hbar v_F}}\Big)\right]
\nonumber \\&&
+\Big(\frac{d}{L}\Big)^{\beta}\bigg[\tilde{t}_{AB}^{(1)}
\cos \Big({\frac{{\bf g} \mu_B \delta B L}{2\hbar v_F}}\Big)
+\tilde{t}_{AB}^{(2)}
\cos \Big({\frac{{\bf g} \mu_B B_T L}{2\hbar v_F}}\Big)\bigg]\bigg\},
\label{Ljcurr}
\end{eqnarray}
where $\delta B=B_A-B_B$, $B_T=B_A+B_B$ and $\beta=(g+1/g)/4-1/2$. Also
$d$ is the diameter of the tube and is of the order of a few
nanometers, ${\bf g}$ is the Land\' e factor, and $\mu_B$ is the
Bohr magneton.  The dimensionless constants $\tilde{t}_{ii}^{(1/2)}$,
$\tilde{t}_{AB}^{(1/2)}$ are proportional to the (small) bare transmission
probabilities of Cooper pairs from the superconductors to the
nanotubes in the unsplit and split processes, respectively.  
The index '$1$' refers to injection of pairs of electrons 
with the same chirality (two right movers or two left movers) into
the wires, while the index '$2$' refers to injection of pairs consisting
of a left-moving and a right-moving electron.
Notice the
highly anomalous length dependence compared to the case of
non-interacting wires, where the Josephson current is inversely
proportional to the separation length $L$.  Because they have the largest 
power-law prefactor, the pair-splitting ($\tilde{t}_{AB}^{(1/2)}$) terms are
enhanced relative to the next largest
contribution by a factor of $(L/d)^{0.6} \gtrsim 63$
for typical nanotube parameters.
Strikingly, as a function of either one of the applied magnetic fields $B_A$ 
and $B_B$, contributions from Cooper pair split processes oscillate with
{\sl half} the frequency of those generated by
unsplit pairs ($\tilde{t}_{ii}^{(2)}$).  We estimate 
the period of these magnetic field oscillations to be in the Tesla
range.  In an actual experiment, in which some flux between the
nanotubes is inevitable, the above critical current will give
the {\sl envelope} for much faster Aharanov-Bohm oscillations (on the
scale of a few Gauss) in the
critical current, but the two types of oscillations can be easily
distinguished by their very different periodicities.  

While the proposed measurements establish that charge
enters from the superconductor in the form of separated electrons with
spin $1/2$ each, they do not establish that these electrons are
actually in the (maximally) entangled singlet state
$[|\uparrow>_{A} |\downarrow>_{B}-|\downarrow>_{A}
|\uparrow>_{B}]/\sqrt{2}$.  This entanglement is encoded in correlations
between the injected spins.  As a
simple but revealing example, consider the joint probability
$P_{\uparrow\uparrow}$ that both electrons in a given pair have spin
``up'' along a selected axis $\hat{\bf z}$.  In the singlet case,
illustrating the EPR 'paradox', once spin up is measured in tube A,
spin down is automatically selected in tube B, and the probability is
zero. Contrast this with the measurement of an unentangled up-down pair
$|\uparrow>_{A\hat{\bf m}}|\downarrow>_{B\hat{\bf m}}$ along an
arbitrary direction $\hat{\bf m}$.  If one attempts to preserve
spin-rotational invariance on average by choosing the axis $\hat{\bf
  m}$ with uniform probability on the unit sphere, the probability
$P_{\uparrow\uparrow}>0$ is non-vanishing, due to pairs that are not
oriented along $\hat{\bf z}$.

We now propose a specific measurement to test the presence of
entanglement through current correlations, i.e. noise.  Specifically,
in the transconductance set-up of Fig.1, what is required
is an experimental measurement of the currents $I_{i\hat{\bf n}}$ of
electrons with a given spin orientation (along $\hat{\bf n}$) in each
nanotube, $i=A,B$.  Experimentally, this could be accomplished by a
variety of spin-filtering techniques, e.g. by attaching two oppositely
polarized half-metallic ferromagnets via {\sl ideal adiabatic
  contacts} to each nanotube (many other schemes, e.g Ref.\cite{spinfilter}
are possible.)  Consider measuring
spin-filtered currents along the ${\bf \hat{z}}$ axis in tube A and
along an axis ${\bf \hat{n}}$ oriented at an angle $\theta$ with
respect to ${\bf \hat{z}}$ in tube B.  When a finite voltage drop
$V_A$ is applied across tube A, the most revealing measurements are
those with no voltage drop across tube B, since then all the current
in it is due to pair splitting processes.

In a manner similar to the one used to compute the conductance matrix for the
system described in Fig.1 we make use of a perturbative Keldysh approach
\cite{Keldysh1,Keldysh2}
to derive forms for the spin-filtered currents and current-current
correlations.
For the case of singlet Cooper pair tunneling, 
the spin-filtered current correlations at zero temperature are found to be:
\begin{eqnarray}
  \langle I_{A \pm \bf \hat{z}} I_{B \pm
    \hat{\bf n}}\rangle &&=
  e \sin^2\frac{\theta}{2} \langle I_{B \hat{\bf n}}\rangle,
  \nonumber \\
  \langle I_{A \pm \bf \hat{z}} I_{B \mp
    \hat{\bf n}}\rangle &&=e
  \cos^2\frac{\theta}{2} \langle I_{B \hat{\bf n}}\rangle,
  \nonumber \\
  \langle I_{B \hat{\bf n}}\rangle &&= \langle I_{B -\hat{\bf
      n}}\rangle,
\label{spincor}
\end{eqnarray}
where $\theta$ is the angle between the ${\bf \hat{n}}$ and ${\bf
  \hat{z}}$ axes. 
From Eq. (\ref{spincor}) we see that when both measurements are made along
the $\hat{\bf z}$-axis ($\theta=0$), as expected, correlations between
like spin currents in the two wires vanish.  Also, measurements on B
can be made in the x-y plane ($\theta=\pi/2$).  Then, as seen in
Eq. (\ref{spincor}), measuring spin-up(or spin-down) in tube A, provides
a 50-percent chance of spin-up and a 50-percent chance of spin-down
for tube B. By contrast, in the case considered above of tunneling of 
classically random unentangled spin-up spin-down pairs, one would expect
zero temperature correlations of the form
\begin{eqnarray}
\langle I_{A \pm \bf \hat{z}} I_{B \pm
\hat{\bf n}}\rangle &&=
\frac{e}{3} [1+\sin^2\frac{\theta}{2}] \langle I_{B \hat{\bf n}}\rangle,
\nonumber \\
\langle I_{A \pm \bf \hat{z}} I_{B \mp
\hat{\bf n}}\rangle &&=
\frac{e}{3} [1+\cos^2\frac{\theta}{2}] \langle I_{B \hat{\bf n}}\rangle.
\end{eqnarray}
Though the correlations show an angular dependence on $\theta$, their
form is very different from the entangled case of Eq. (\ref{spincor}).
Specifically, as anticipated for measurements along the $\hat{\bf z}$
axis in both tubes ($\theta=0$), we have $<I_{A \pm\bf \hat{z}} I_{B
  \pm \hat{\bf z}}> =e/3 <I_{B \hat{\bf z}}>$, which is
non-zero, in stark contrast to the entangled case.

To conclude, in the vast search for physical realizations of
entanglement, we have described one method of extracting singlet pairs
from a superconducting source. If employing nanotubes for this purpose
indeed proves tractable, the next stage in the realm of quantum
information would involve new challenges such as probing information
at the single electron level and building arrays of coupled logic
gates.  In the fields of nanoscience and Luttinger physics, attention,
both theoretical and experimental, has fallen on bringing effectively
one-dimensional systems into contact with superconductors
\cite{Jos1}.  
Here we have hoped to provide
more food for thought in these fields by describing two nanotubes in
contact with a gapped BCS superconductor.

While finalizing this work we became aware of a related independent 
proposal by  P. Recher and D. Loss \cite{recher}. The analysis  
of \cite{recher} is based on a set-up similar to the one we describe here,
yielding similar results with the ones we derive in the first part of our
paper.

 We are grateful to Daniel Loss for discussions.
This research is supported by NSF grants DMR-9985255, DMR-97-04005,
DMR95-28578, PHY94-07194, and the Sloan and Packard foundations.

\end{multicols}

\begin{thebibliography}{99}
\bibitem{EPR} A. Einstein, B. Podolski, N. Rosen, Phys. Rev. {\bf 47},
777 (1935).
\bibitem{entangle1} A. Aspect, J. Dalibard, C. Roger, Phys. Rev. Lett
{\bf 49}, 1804 (1982); D. Jaksch, H. J. Briegel, 
J. I. Cirac, C. W. Gardiner, P. Zoller, 
Phys. Rev. Lett. {\bf 82}, 1975 (1999).
\bibitem{loss} P. Recher, E. V. Sukhorukov, D. Loss, Phys. Rev. B {\bf 63}, 
165314 (2001); G. B. Lesovik, T. Martin, G. Blatter, cond-mat/0009193.
\bibitem{BalentsEgger} L. Balents and R. Egger, Phys. Rev. B {\bf 64}, 
  35310 (2001).
\bibitem{alphenaar} K. Tsukagoshi, B. W. Alphenaar, H. Ago, Nature {\bf 401},
571 (1999).
\bibitem{dekker} A. Bachtold, P. Hadley, T. Nakanishi, C. Dekker, Science
{\bf 294}, 1317 (2001); 
P. G. Collins, M. S. Arnold, P. Avouris, Science {\bf 297}
,706 (2001).
\bibitem{nanexp2} M. Bockrath, {\it et al.}, Nature {\bf 397}, 598 (1999);
Z. Yao, H. Postma, L. Balents, C. Dekker,
Nature {\bf 402}, 273 (1999); H. Postma, M. de Jonge, Z. Yao, C. Dekker, 
Phys. Rev. B {\bf 62}, 10653 (2000). 
\bibitem{Klm} C. L. Kane, L. Balents, M. P. A. Fisher, Phys. Rev. Lett. 
{\bf 79}, 5086 (1997).
\bibitem{egger} R. Egger, A. Gogolin, Phys. Rev. Lett. {\bf 79},
5082 (1997).
\bibitem{Keldysh1} L. V. Keldysh, ZhETF {\bf 47}, 1515 (1964) [Sov.
Phys. JETP {\bf 20}, 1018 (1965)]; M. P. A. Fisher and W. Zwerger,
Phys. Rev. B {\bf 32}, 6190 (1985).
\bibitem{Keldysh2} C. L. Kane and M. P. A. Fisher, Phys. Rev. Lett.
{\bf 72}, 724 (1994).
\bibitem{spinfilter} P. Recher, E. V. Sukhorukov, D. Loss, 
Phys. Rev. Lett. {\bf 85}, 1962 (2000).
\bibitem{Jos1} A. F. Morpurgo, J. Kong, C. M. Marcus,
H. Dai, Science {\bf 286}, 263 (1999); 
A. Yu. Kasumov {\it et al.}, Science {\bf 284}, 1508 (1999)
\bibitem{Jos2} I. Affleck, J.-S. Caux, A. M. Zagoskin, 
Phys. Rev. B, {\bf 62}, 1433 (2000); R. Fazio, F. W. J. Hekking, 
A. A. Odintsov, Phys. Rev. Lett {\bf 74}, 1843 (1995);
D. L. Maslov, M. Stone, P. M. Goldbart and 
D. Loss, Phys. Rev. B {\bf 53}, 1548 (1996); Y. Takane, J. Phys. Soc. Japan, 
{\bf 66}, 537 (1997). 
\bibitem{recher} P. Recher and D. Loss, cond-mat/0112298.
\end{thebibliography}
\end{document}